New Series of Nickel-Based Pnictide Oxide Superconductors ($Ni_2Pn_2$)($Sr_4Sc_2O_6$) (*Pn* = P, As)

Yutaka Matsumura[1], Hiraku Ogino[1,2*], Shigeru Horii[2,3], Yukari Katsura[1], Kohji Kishio[1,2], and Jun-ichi Shimoyama[1,2]

[1]Department of Applied Chemistry, The University of Tokyo, 7-3-1 Hongo, Bunkyo-ku, Tokyo 113-8656, Japan

[2]JST-TRIP, Sanban-cho, Chiyoda-ku, Tokyo 102-0075, Japan

[3]Department of Environmental Systems Engineering, Kochi University of Technology, Kami-shi, Kochi 782-8502, Japan

We have discovered new nickel-based pnictide oxide superconductors, ($Ni_2Pn_2$)($Sr_4Sc_2O_6$) (*Pn* = P, As). These compounds have a tetragonal unit cell with a space group of *P*4/*nmm* and they consist of alternate stacking of anti-fluorite $Ni_2Pn_2$ layers and $K_2NiF_4$-type $Sr_4Sc_2O_6$ blocking layers. Lattice parameters were $a$ = 4.044 Å and $c$ = 15.23 Å for ($Ni_2P_2$)($Sr_4Sc_2O_6$) and $a$ = 4.078 Å and $c$ = 15.41 Å for ($Ni_2As_2$)($Sr_4Sc_2O_6$), indicating their thicker blocking layers than that of LaNiPO ($c$ ~ 8.1 Å). Both ($Ni_2P_2$)($Sr_4Sc_2O_6$) and ($Ni_2As_2$)($Sr_4Sc_2O_6$) exhibited superconductivity with zero resistivity at 3.3 K and 2.7 K, respectively. The perfect diamagnetism observed in both compounds guaranteed their bulk superconductivity.

*E-mail address: tuogino@mail.ecc.u-tokyo.ac.jp

The discovery of high-$T_c$ superconductivity in LaFeAs(O,F)[1] has triggered a new superconducting vein of layered pnictide oxide systems. So far, several group of superconductors have been discovered in this system, such as *RE*Fe*Pn*O(abbreviated as 1111, *RE* = rare earth elements, *Pn* = P, As)[2], *AE*Fe$_2$*Pn*$_2$(abbreviated as 122, *AE* = alkali earth elements)[3], LiFeAs[4], and Fe*Ch*(*Ch* = charcogenides)[5]. All the materials have layered structure with alternate stacking of anti-fluorite Fe*Pn* layers and blocking layers, such as fluorite *RE*O layers or *AE* sheets. On the other hand, transition metal pnictide oxides having the ZrThCrSi structure, such as REMn*Pn*O[6], *RE*Co*Pn*O[7], and *RE*Ni*Pn*O[8], have been already reported and, similarly, various transition metal pnictides with $ThCr_2Si_2$ structure are known. Among them, compounds containing anti-fluorite NiP layers show superconductivity with relatively low $T_c$'s, such as 4 K for LaNiPO, 7 K for LaNiP(O,F), and 3 K for $BaNi_2P_2$[9].

Recently we have discovered a new iron phosphide oxide superconductor $(Fe_2P_2)(Sr_4Sc_2O_6)$[10] with $T_c$~17 K. This material has similar structure with recently discovered $(Fe_2As_2)(Sr_3Sc_2O_5)$[11] and a layer stacking combination of anti-fluorite FeP layers and perovskite-type $Sr_2ScO_3$ layers. Consecutive discoveries of isostructural compounds, such as $(Fe_2As_2)(Sr_4Sc_2O_6)$[12], $(Fe_2As_2)(Sr_4Cr_2O_6)$[12] and $(Co_2As_2)(Sr_4Sc_2O_6)$[13], indicated large variety of constituent elements in this system.

In the present study, nickel-based pnictide oxides $(Ni_2P_2)(Sr_4Sc_2O_6)$ and $(Ni_2As_2)(Sr_4Sc_2O_6)$ have been successfully synthesized and their physical properties were characterized. Both materials showed bulk superconductivity with $T_c$ = 3.3 K $(Ni_2P_2)(Sr_4Sc_2O_6)$ and 2.7 K $(Ni_2As_2)(Sr_4Sc_2O_6)$.

Polycrystalline samples with nominal compositions of $(Ni_2P_2)(Sr_4Sc_2O_6)$ and $(Ni_2As_2)(Sr_4Sc_2O_6)$ were synthesized by solid state reaction starting from SrO(2N), Sr(2N), $Sc_2O_3$(4N) and Ni(3N) and P(2N) or NiAs(3N). Appropriate amounts of starting reagents were mixed and pelletized in a dry argon atmosphere. Sintering was done at 1100 - 1200°C for 18-48 h in evacuated quartz ampoules.

Phase identification and determination of lattice constants were performed by X-ray diffraction

(XRD) using a RIGAKU Ultima-IV diffractometer. The intensity data were collected in the $2\theta$ range of 5–80° at a step of 0.02° (Cu-$K_\alpha$) using silicon powder as the internal standard. High-resolution images were taken by JEOL JEM-2010F field emission Transmission Electron Microscopy (TEM). Magnetic susceptibility measurement was performed by Superconducting Quantum Interference Device (SQUID) magnetometer (Quantum Design MPMS-XL5s). Electric resistivity was measured by the AC four-point probe method using Quantum Design PPMS under fields up to 2 T.

Powder XRD analysis revealed that both $(Ni_2P_2)(Sr_4Sc_2O_6)$ and $(Ni_2As_2)(Sr_4Sc_2O_6)$ were formed as main phases by sintering above 1100°C. Figure 1 shows XRD patterns of $(Ni_2P_2)(Sr_4Sc_2O_6)$ synthesized at 1100°C for 48 h and $(Ni_2As_2)(Sr_4Sc_2O_6)$ synthesized at 1200°C for 40 h together with simulated patterns. Most of the peaks observed were assigned as peaks due to $(Ni_2P_2)(Sr_4Sc_2O_6)$ or $(Ni_2As_2)(Sr_4Sc_2O_6)$ except small amount of nonsuperconducting impurities SrP and $Sc_2O_3$. Intensities of impurites were less than 5% of the main phase. Lattice constants of $(Ni_2P_2)(Sr_4Sc_2O_6)$ were calculated to be $a$ = 4.044(3) Å and $c$ = 15.23(2) Å and its cell volume $V_{cell}$ was 249.09 Å$^3$. Slightly large unit cell was confirmed for $(Ni_2As_2)(Sr_4Sc_2O_6)$ with $a$ = 4.078(7) Å, $c$ = 15.41(4) Å and $V_{cell}$ = 256.19 Å$^3$, reflecting larger atomic size of arsenic than that of phosphorous. It was found that both compounds have long $a$-axis, short $c$-axis and small $V_{cell}$ compared to those of iron based relatives ($a$ = 4.016 Å, $c$ = 15.543 Å and $V_{cell}$ =250.68 Å$^3$ for $(Fe_2P_2)(Sr_4Sc_2O_6)$ and $a$ = 4.049 Å, $c$ = 15.809 Å and $V_{cell}$ =259.18 Å$^3$ for $(Fe_2As_2)(Sr_4Sc_2O_6)$), while ionic radius of $Ni^{2+}$ is larger than $Fe^{2+}$(low spin). Similar tendency is also confirmed in the 1111 compounds La*MPn*O (*M* = Fe, Ni: *Pn* = P, As) through comparisons of their reported lattice constants and $V_{cell}$'s. These suggest that the Ni*Pn* tetrahedra in the layered pnictide oxides tend to have low symmetry expanding along the Ni*Pn* layer and shrinking along the $c$-axis, while more symmetric tetrahedra are formed at the Fe*Pn* layers. Although low phase purity prevented us from determining atomic positions of $(Ni_2P_2)(Sr_4Sc_2O_6)$ and $(Ni_2As_2)(Sr_4Sc_2O_6)$ at the present stage, further investigation on the structural analyses will reveal an intrinsic character of the Ni*Pn* layers.

The zero-field-cooled (ZFC) and field-cooled (FC) magnetization curves for sintered bulk samples of $(Ni_2P_2)(Sr_4Sc_2O_6)$ and $(Ni_2As_2)(Sr_4Sc_2O_6)$ are shown in Fig. 2. Applied external fields were 1 Oe for $(Ni_2P_2)(Sr_4Sc_2O_6)$ and 0.1 Oe for $(Ni_2As_2)(Sr_4Sc_2O_6)$. Diamagnetisms were observed below 3.2 K for $(Ni_2P_2)(Sr_4Sc_2O_6)$ and 2.5 K for $(Ni_2As_2)(Sr_4Sc_2O_6)$. Superconducting volume fractions of $(Ni_2P_2)(Sr_4Sc_2O_6)$ and $(Ni_2As_2)(Sr_4Sc_2O_6)$ estimated from ZFC magnetization at 2 K were 140 and 150%, respectively, indicating their bulk superconductivity. Because the amount of impurity phases was not so high, the observed larger diamagnetism than the ideal perfect diamagnetism is explained by the demagnetization effect corresponding to the sample shape.

Temperature dependences of resistivity under several magnetic fields for $(Ni_2P_2)(Sr_4Sc_2O_6)$ and $(Ni_2As_2)(Sr_4Sc_2O_6)$ are shown in Figs. 3(a) and 3(b), respectively. Insets show their resistivity curves measured under 0 Oe up to 300 K. Both $(Ni_2P_2)(Sr_4Sc_2O_6)$ and $(Ni_2As_2)(Sr_4Sc_2O_6)$ exhibited metallic behavior in the normal state. Relatively high resistivity and high residual resistivity of $(Ni_2P_2)(Sr_4Sc_2O_6)$ are probably due to coexisting impurity phases. Sharp resistivity drops were observed for both compounds. $T_{c(onset)}$'s and $T_{c(\rho=0)}$'s are 3.6 and 3.3 K for $(Ni_2P_2)(Sr_4Sc_2O_6)$ and 2.9 and 2.7 K for $(Ni_2As_2)(Sr_4Sc_2O_6)$ in 0 T. Their superconducting transitions became broader with an increase of magnetic field as in the case of $(Fe_2P_2)(Sr_4Sc_2O_6)$[10]. The Upper critical fields defined by the temperatures where $\rho(T) = 0.9\rho_{normal}$ of the $(Ni_2P_2)(Sr_4Sc_2O_6)$ and $(Ni_2As_2)(Sr_4Sc_2O_6)$ were estimated to be ~1.2 T and ~0.5 T at 2 K, respectively.

The newly found compounds $(Ni_2P_2)(Sr_4Sc_2O_6)$ and $(Ni_2As_2)(Sr_4Sc_2O_6)$ belongs to the same structural family of a recently found layered pnictide oxides $(Fe_2P_2)(Sr_4Sc_2O_6)$[10] and $(Fe_2As_2)(Sr_4Sc_2O_6)$[12]. The $T_c$'s of La*MPn*O and $(M_2Pn_2)(Sr_4Sc_2O_6)$ systems ($M$ = Fe, Ni: $Pn$ = P, As) are summarized in Table I. One can find some similarities between La*MPn*O and $(M_2Pn_2)(Sr_4Sc_2O_6)$ systems. At first, both systems show superconductivity except a combination of $M$ = Fe and $Pn$ = As without intentional carrier doping. The $T_c$'s for nickel based compounds are comparable and relatively low. A large difference in $T_c$ between $(Fe_2P_2)(Sr_4Sc_2O_6)$ and LaFePO is partly explained by the distortion of $FeP_4$ tetrahedra. The former compound has more symmetric $FeP_4$ tetrahedra. For detailed comparison between LaNi*Pn*O and $(Ni_2Pn_2)(Sr_4Sc_2O_6)$, determinations of atomic positions of $(Ni_2Pn_2)(Sr_4Sc_2O_6)$ are needed using high quality samples.

Discoveries of the two compounds with Ni*Pn* layer in the present study and recently reported $(Co_2As_2)(Sr_4Sc_2O_6)$ suggest that further developments of new materials from $(M_2Pn_2)(Sr_4Sc_2O_6)$ ($M$ = Mn to Zn, $Pn$ = P to Bi) are promising. Combined with the recent discovery of a chromium containing iron pnictide oxide $(Fe_2As_2)(Sr_4Cr_2O_6)$[11)], a large increase in number of the layered pnictide oxides having the perovskite-type blocking layer can be expected.

In summary, new layered nickel pnictide oxides, $(Ni_2P_2)(Sr_4Sc_2O_6)$ and $(Ni_2As_2)(Sr_4Sc_2O_6)$, have been successfully synthesized and their physical properties were characterized. These compounds have the same layer stacking structure of anti-fluorite Ni*Pn* and perovskite-type $Sr_2ScO_3$ layers. In both resistivity and magnetization measurements, they showed superconductivity with $T_{c(\rho=0)}$'s of 3.3 K for $(Ni_2P_2)(Sr_4Sc_2O_6)$ and 2.7 K for $(Ni_2As_2)(Sr_4Sc_2O_6)$. Similar to the LaNi*Pn*O compounds ($Pn$ = P, As), the $(Ni_2Pn_2)(Sr_4Sc_2O_6)$ showed superconductivity without intentional doping. This fact suggests layered pnictide oxides with perovskite-type oxide layer have similar electronic structure with *REMPn*O. Therefore, this system will be a new family of the layered pnictide oxide superconductors.

Acknowledgements

This work was partly supported by Grant-in-Aid for Young Scientists (B) no. 21750187, 2009, supported by the Ministry of Education, Culture, Science and Technology (MEXT).

## References

1) Y. Kamihara, T. Watanabe, M. Hirano, and H. Hosono: J. Am. Chem. Soc. **130** (2008) 3296.

2) P. Quebe, L. Terbu¨chte, and W. Jeitschko: J. Alloys Compd. **302** (2000) 70.

3) M. Rotter, M Tegel, and D. Johrendt: Phys. Rev. Lett. **101** (2008) 107006.

4) M. Pitcher, D. Parker, P. Adamson, S. Herkelrath, A. Boothroyd, R. Ibberson, M. Brunelli, and S. Clarke: Chem. Commun. (2008) 5918.

5) F. C. Hsu, J. Y. Luo, K. W. The, T. K. Chen, T. W. Huang, P. M. Wu, Y. C. Lee, Y. L. Huang, Y. Y. Chu, D. C. Yan, and M. K. Wu: Proc. Natl. Acad. Sci. U.S.A. **105** (2008) 14262.

6) A.T. Nientiedt, W. Jeitschko, P.G. Pollmeier, and M. Brylak: Zeitschrift fuer Naturforschung B **52** (1997) 560.

7) B. I. Zimmer, W. Jeitschko, J. H. Albering, R. Glaum, M. Reehuis: J. Alloys Compd. **229** (1995) 238.

8) T. Watanabe, H. Yanagi, T. Kamiya, Y. Kamihara, H. Hiramatsu, M. Hirano, and H. Hosono: Inorg. Chem. **46** (2007) 7719.

9) T. Mine, H. Yanagi, T. Kamiya, Y. Kamihara, M. Hirano, and H. Hosono: Solid State Commun. **147** (2008) 111.

10) H. Ogino, Y. Matsumura, Y. Katsura, K. Ushiyama, S. Horii, K. Kishio, and J. Shimoyama: arXiv: 0903.3314(unpublished)

11) X. Zhu, F. Han, G. Mu, B. Zeng, P. Cheng, B. Shen, H.H. Wen: Phys. Rev. B 79 (2009) 024516.

12) H. Ogino, Y. Katsura, S. Horii, K. Kishio, and J. Shimoyama: arXiv: 0903.5124(unpublished)

13) Y. L. Xie, R. H. Liu, T. Wu, G. Wu, Y. A. Song, D. Tan, X. F. Wang, H. Chen, J. J. Ying, Y. J. Yan, Q. J. Li, X. H. Chen: arXiv:0903.5484(unpublished)

14) Y. Kamihara, H. Hiramatsu, M. Hirano, R. Kamihara, H. Yanagi, T. Kamiya, and H. Hosono: J. Am. Chem. Soc. **128** (2006) 10012.

15) Z. Li, G. F. Chen, J. Dong, G. Li, W. Z. Hu, J. Zhou, D. Wu, S. K. Su, P. Zheng, N. L. Wang, and J. L. Luo: Phys. Rev. B **78** (2008) 060504.

**Figure captions**

Figure 1 Observed and calculated powder XRD patterns of $(Ni_2P_2)(Sr_4Sc_2O_6)$ (a) and $(Ni_2As_2)(Sr_4Sc_2O_6)$ (b).

Figure 2 Temperature dependence of ZFC and FC magnetization curves of $(Ni_2P_2)(Sr_4Sc_2O_6)$ (open circle) and $(Ni_2As_2)(Sr_4Sc_2O_6)$ (closed circle) measured under 1 and 0.1 Oe, respectively.

Figure 3 Temperature dependences of resistivity of $(Ni_2P_2)(Sr_4Sc_2O_6)$ (a) and $(Ni_2As_2)(Sr_4Sc_2O_6)$ (b) in 0, 0.2, 0.5, 1, and 2 T. Temperature dependences of resistivity under 0 T from 2 to 300 K are shown in the insets.

Table I $T_c$'s for La*MPn*O and $(M_2Pn_2)(Sr_4Sc_2O_6)$ (*M* = Fe, Ni: *Pn* = P, As).

| | La*MPn*O | | $(TM_2Pn_2)(Sr_4Sc_2O_6)$ | |
|---|---|---|---|---|
| | Fe | Ni | Fe | Ni |
| P | 4 K (undoped)[14)]<br>7 K (F-doped)[14)] | 4 K[8)] | ~ 17 K (undoped)[10)] | 3.3 K* |
| As | non super (undoped)[1)]<br>~ 26 K (F-doped)[1)] | 2.75 K[15)] | non super (undoped)[12)] | 2.7 K* |

* this work

Figure 1

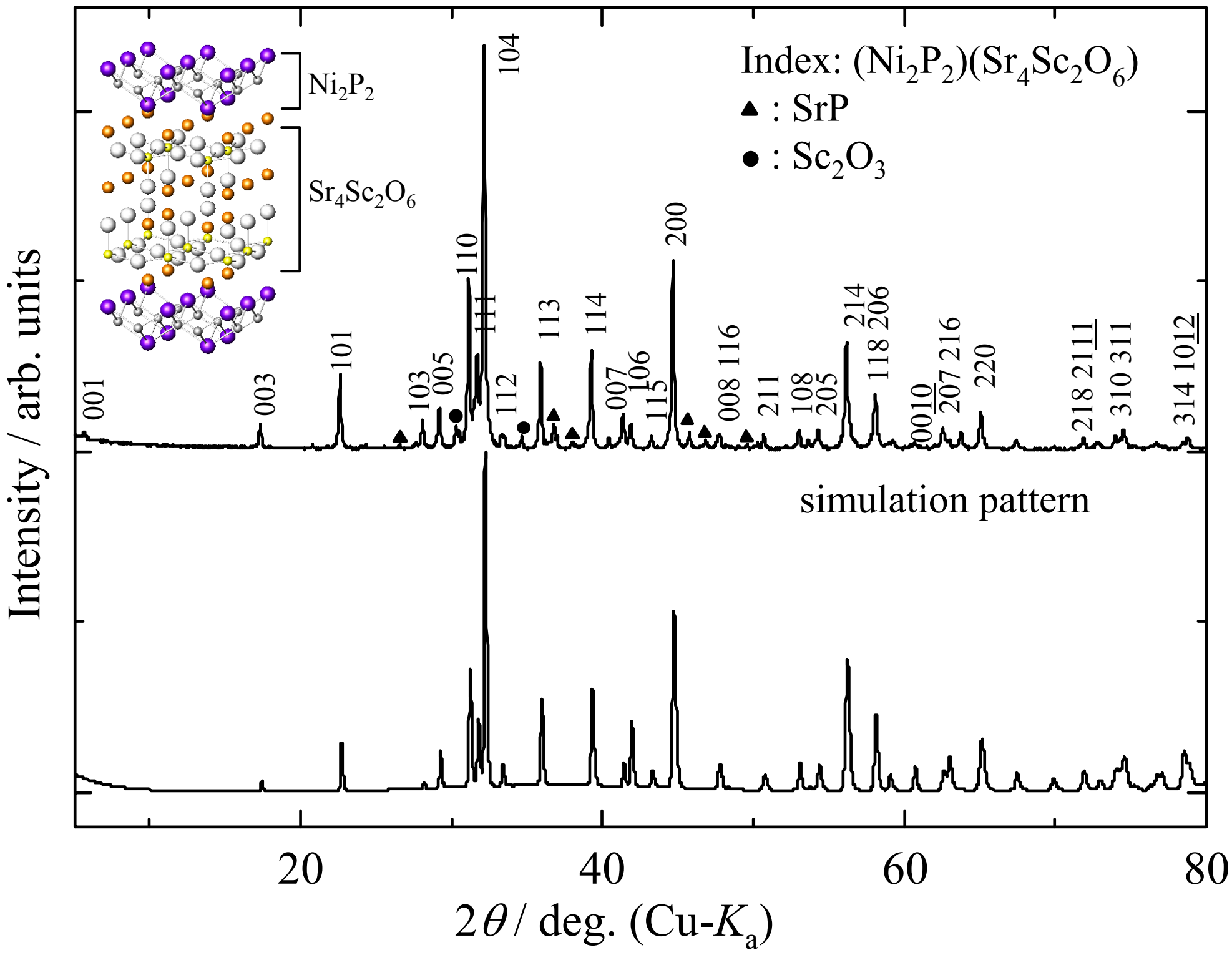

(a)

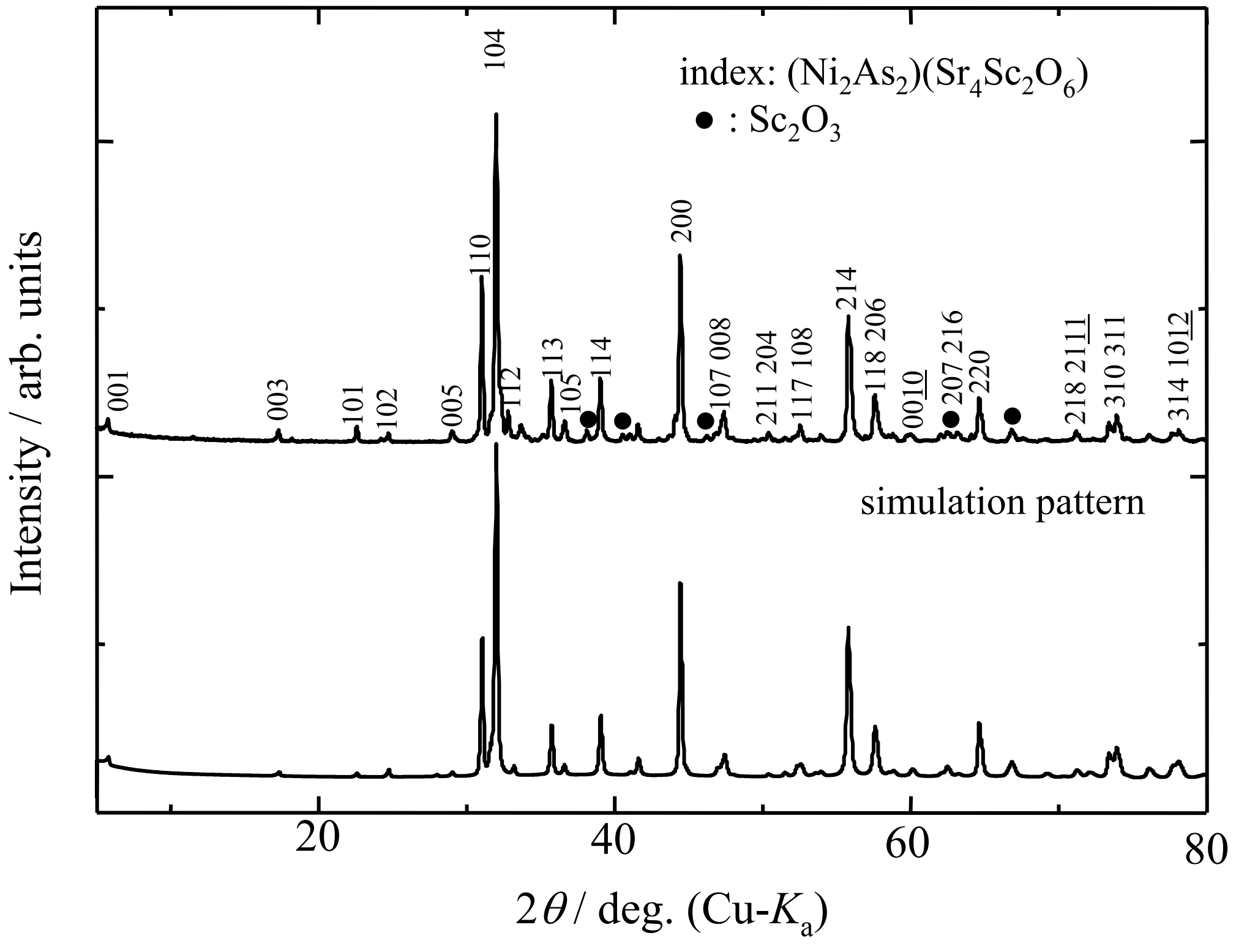

(b)

Figure 2

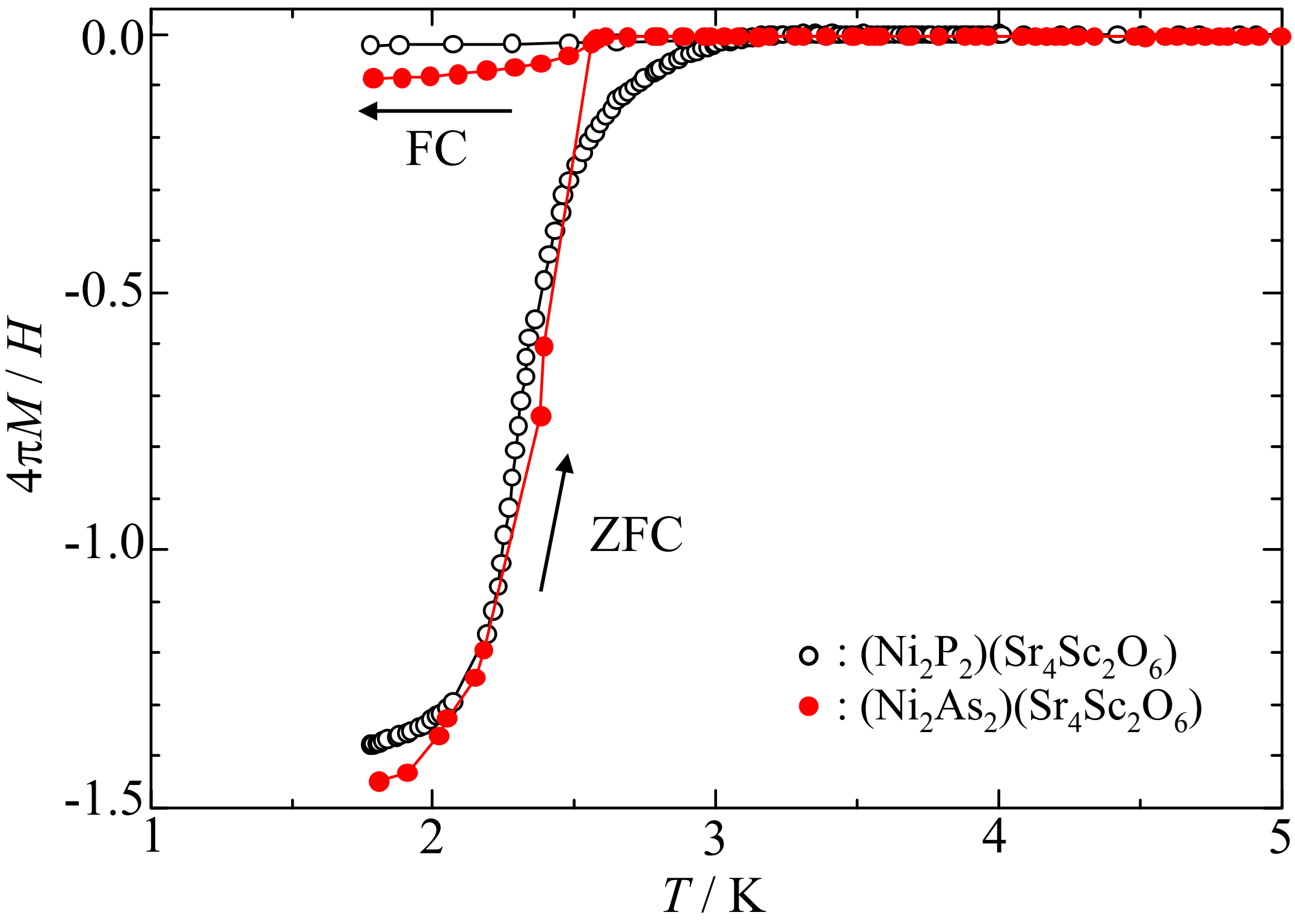

Figure 3

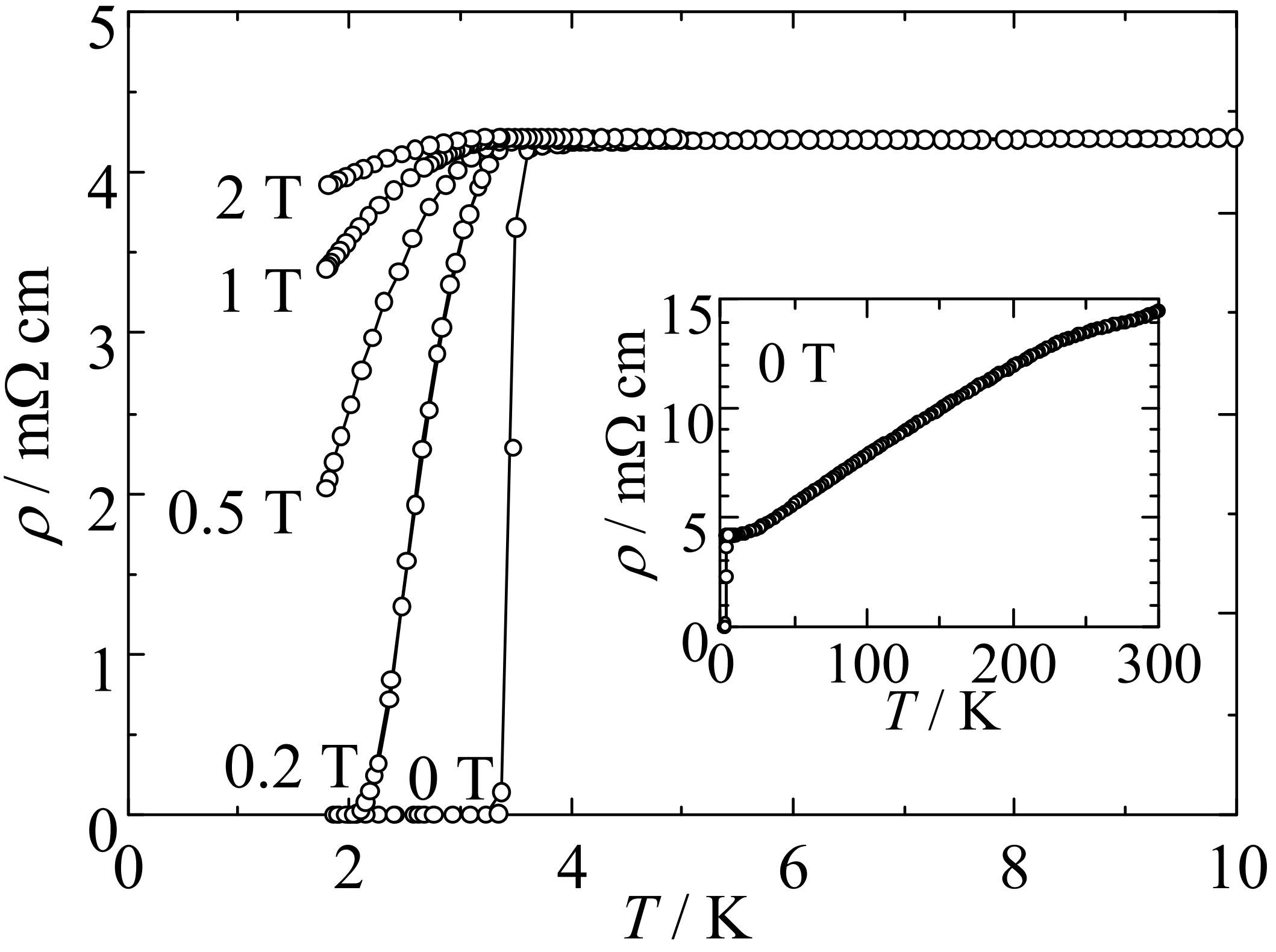

(a)

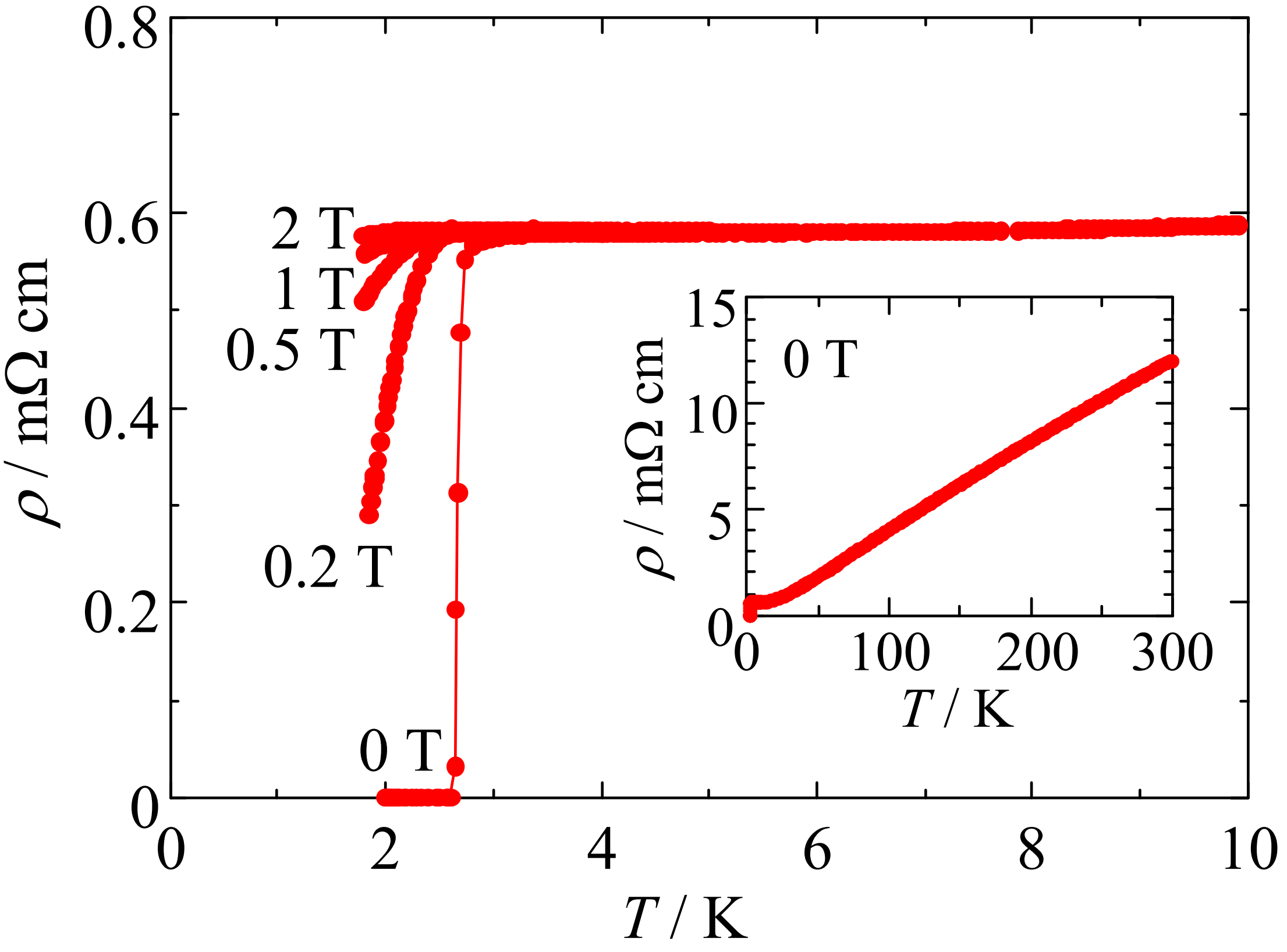

(b)